\documentclass[aps,pre,showpacs,noshowkeys,amsmath,amssymb,amsfonts,superscriptaddress,longbibliography,reprint]{revtex4-1}
\usepackage[english]{babel}

\usepackage{graphicx}
\usepackage{bm}
\usepackage{physics}
\usepackage{mathtools}
\usepackage{gensymb}
\usepackage[shortlabels]{enumitem}

\newcommand{\V}{\mathcal{V}}

\setcitestyle{super}
\usepackage{caption}
\usepackage{subcaption}
\DeclareCaptionLabelSeparator{bar}{~\rule[-0.4ex]{0.2ex}{1em}~}
\DeclareCaptionLabelFormat{subfor}{\textbf{#2}}
\newcommand*\bfcaption[2]{\caption[#1]{\textbf{#1.}#2}}
\usepackage{xcolor}
\definecolor{UBcolor}{HTML}{007CC1}
\usepackage[colorlinks=true,pdfnewwindow=true,linkcolor=UBcolor,citecolor=UBcolor,urlcolor=UBcolor,breaklinks=true,linktocpage]{hyperref}
\usepackage[all]{hypcap}
\usepackage[nameinlink,capitalise]{cleveref}
\crefname{SI section}{SI Section}{SI Sections}
\Crefname{SI section}{SI Section}{SI Sections}
\begin{document}

\title{Cellular Sensing Governs the Stability of Chemotactic Fronts}

\author{Ricard Alert}
%\altaffiliation{These authors contributed equally to this work.}
\email{ricard.alert@princeton.edu}
\affiliation{Princeton Center for Theoretical Science, Princeton University, Princeton, NJ 08544, USA}
\affiliation{Lewis-Sigler Institute for Integrative Genomics, Princeton University, Princeton, NJ 08544, USA}
\affiliation{Max Planck Institute for the Physics of Complex Systems, N\"{o}thnitzerst. 38, 01187 Dresden, Germany}
\affiliation{Center for Systems Biology Dresden, Pfotenhauerst. 108, 01307 Dresden, Germany}

\author{Alejandro Mart\'{i}nez-Calvo}
\affiliation{Princeton Center for Theoretical Science, Princeton University, Princeton, NJ 08544, USA}
\affiliation{Department of Chemical and Biological Engineering, Princeton University, Princeton, NJ 08544, USA}

\author{Sujit S. Datta}
\email{ssdatta@princeton.edu}
\affiliation{Department of Chemical and Biological Engineering, Princeton University, Princeton, NJ 08544, USA}

\date{\today}

\begin{abstract}
In contexts ranging from embryonic development to bacterial ecology, cell populations migrate chemotactically along self-generated chemical gradients, often forming a propagating front. Here, we theoretically show that the stability of such chemotactic fronts to morphological perturbations is determined by limitations in the ability of individual cells to sense and thereby respond to the chemical gradient. Specifically, cells at bulging parts of a front are exposed to a smaller gradient, which slows them down and promotes stability, but they also respond more strongly to the gradient, which speeds them up and promotes instability. We predict that this competition leads to chemotactic fingering when sensing is limited at too low chemical concentrations. Guided by this finding and by experimental data on \textit{E. coli} chemotaxis, we suggest that the cells' sensory machinery might have evolved to avoid these limitations and ensure stable front propagation. Finally, as sensing of any stimuli is necessarily limited in living and active matter in general, the principle of sensing-induced stability may operate in other types of directed migration such as durotaxis, electrotaxis, and phototaxis.
\end{abstract}

\maketitle

Fronts are propagating interfaces that allow one spatial domain to invade another. They are ubiquitous in nature, arising for example during phase transitions, autocatalytic chemical reactions, and flame propagation \cite{Cross1993,Cross2009,Pismen2006,Desai2009a,VanSaarloos2003,Colinet2010}. Biology also abounds with examples, such as fronts of gene expression during development, electric signals in the heart and the brain, infection during disease outbreaks, and expanding populations in ecosystems \cite{Murray2002,Edelstein-Keshet2005,Negrete2021}. These examples can all be modeled as reaction-diffusion systems (e.g., using the Fisher-KPP equation\cite{Fisher1937,Kolmogorov1937}), for which both the motion and morphologies of fronts are well understood \cite{Cross1993,Cross2009,Pismen2006,Desai2009a,VanSaarloos2003,Colinet2010}. 

Another prominent and separate class of fronts is that of \emph{chemotactic fronts}, in which active agents collectively migrate in response to a self-generated chemical gradient. These fronts have long been observed in bacterial populations, enabling cells to escape from harmful conditions, colonize new terrain, and coexist \cite{Murray2003,Colin2021,Adler1966,Budrene1991,Budrene1995,Fu2018,Cremer2019,Gude2020a,Bhattacharjee2021a,Bai2021}. More generally, collective chemotaxis plays crucial roles in slime mold aggregation \cite{Keller1970}, embryonic development \cite{Scarpa2016,Painter1999,Theveneau2010}, immune response \cite{Tweedy2020}, and cancer progression \cite{Malet-Engra2015,Puliafito2015}. Beyond cell populations, enzymes \cite{Jee2018,Agudo-Canalejo2018a,Mohajerani2018} and synthetic active colloids \cite{Illien2017,Liebchen2018,Stark2018} also exhibit collective chemotaxis. Therefore, studies of chemotactic fronts are of broad interest in biological and active matter physics. However, while the motion of chemotactic fronts can be successfully modeled in certain cases \cite{Keller1971,Keller1971a,Brenner1998,Fu2018,Seyrich2019,Cremer2019,Bhattacharjee2021a,Narla2021,Bai2021}, a general understanding of how their morphologies evolve -- akin to that of reaction-diffusion systems -- remains lacking.

For example, a fundamental feature of a front is its morphological stability: Do shape perturbations decay or grow over time? This question is well-studied in non-living systems. In many cases, flat fronts are unstable, leading to striking dendritic patterns at fluid and solid interfaces as in the case of the well-studied Saffman-Taylor and Mullins-Sekerka instabilities \cite{Gollub1999,Gonzalez-Cinca2004,Colinet2010,Casademunt2004,Langer1980,Ben-Jacob1990,Saffman1958,Mullins1964}. In active and living matter, front instabilities underlie fingering patterns in active colloids \cite{Driscoll2017}, growing tumors \cite{Khain2006,Bogdan2018} and bacterial biofilms \cite{Ben-Jacob2000,Allen2019,Mueller2021,Kitsunezaki1997,Muller2002,Farrell2013,Doostmohammadi2016a,Wang2017a,Trinschek2018,Yan2019,Fei2020}, as well as mechanically-competing tissues \cite{Williamson2018,Buscher2020} and spreading epithelia \cite{Alert2020,Perez-Gonzalez2019,Alert2019,Trenado2021}. Front stability has also been analyzed when chemotaxis supplements effects like growth and mechanical interactions \cite{Ben-Jacob1994,Brenner1998,Ben-Jacob2000,BenAmar2013,BenAmar2016a}.

Nevertheless, the conditions for the stability of chemotactic fronts remain unknown. Unlike reaction-diffusion systems, which rely only on scalar couplings between fields, chemotaxis couples the population density to the \emph{gradient} of a chemical signal. Thus, the analytical techniques used to study the stability of reaction-diffusion fronts \cite{Desai2009a} cannot be directly applied to their chemotactic counterparts \cite{Funaki2006}.

Here, through direct analysis of their governing equations, we determine the conditions for the linear stability of chemotactic cell fronts. We find that front stability is determined by the ability of cells to sense chemical stimuli at different concentrations, which modulates their response to the chemical gradient and subsequent propagation speeds at different locations along the front. Our calculations reveal two competing mechanisms governing front stability: When cells move ahead of the front, they absorb chemoattractant, causing follower cells to be exposed to (i) a smaller chemical gradient, which slows cells down and promotes stability, and (ii) a lower chemical concentration, which increases the cellular response, speeds cells up, and promotes instability. We predict a chemotactic fingering instability when sensing is limited at low chemical concentrations, for which the tactic response is strong. Therefore, our work links the properties of the sensory machinery of individual cells to the population-scale morphology of chemotactic fronts. Finally, we suggest that this machinery might have evolved to push sensing limitations to high chemical concentrations in order to ensure stable collective chemotaxis.

\begin{figure}[tb!]
\begin{center}
\includegraphics[width=\columnwidth]{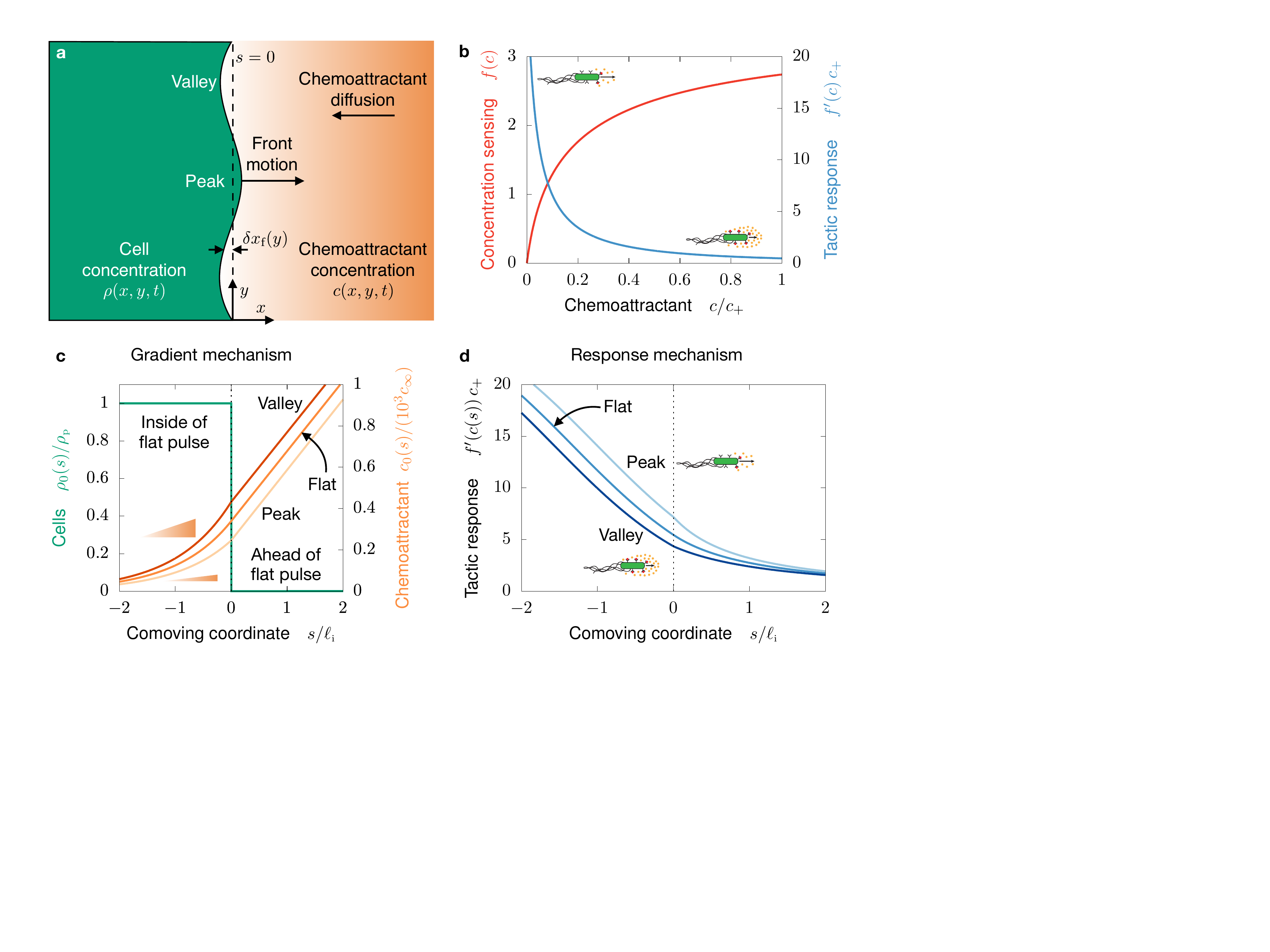}
\end{center}
  {\phantomsubcaption\label{Fig schematic}}
  {\phantomsubcaption\label{Fig sensing-response}}
  {\phantomsubcaption\label{Fig gradient-mechanism}}
  {\phantomsubcaption\label{Fig response-mechanism}}
\bfcaption{Competing mechanisms of chemotactic front stability}{ \subref*{Fig schematic}, Schematic of a cell population (green) moving up a chemoattractant gradient (orange). We analyze the stability of a reference flat front (dashed line, located at the origin of the comoving coordinate $s\equiv x-v_0 t =0$) to perturbations $\delta x_{\text{f}}(y)$, which create peaks and valleys. Note that we did not represent the $y$-dependence of the chemoattractant field. \subref*{Fig sensing-response}, Whereas the ability of cells to sense chemoattractant, $f(c)$, increases with chemoattractant concentration, their tactic response to gradients, $f'(c)$, decreases. \subref*{Fig gradient-mechanism}, Assuming a step profile of cells (green), the chemoattractant profiles for the reference flat front (\cref{eq chemo-flat-solutions}) as well as for peaks and valleys of a perturbed front are shown by the orange curves. The comoving coordinate is rescaled by the internal decay length $\ell_{\text{i}}$ (see text). As depicted in the insets, the chemoattractant gradient at $s=0$ is higher in valleys and lower in peaks, favoring front stability (first term in \cref{eq drift-perturbations-front}). \subref*{Fig response-mechanism}, As depicted in the insets, the cellular response at $s=0$ is stronger in peaks and weaker in valleys, favoring instability (second term in \cref{eq drift-perturbations-front}). In \subref*{Fig gradient-mechanism} and \subref*{Fig response-mechanism}, we used front perturbations $\delta x_{\text{f}}(y) = \delta A \sin(ky)$ with amplitude $\delta A = 2$ $\mu$m and wavelength $\lambda = 2\pi/k = 2$ mm. Parameter values are in \cref{t parameters}.}
\label{Fig1}
\end{figure}

\textbf{Keller-Segel equations.} Following classic work by Keller and Segel \cite{Keller1971,Keller1971a}, we model chemotactic fronts through the coupled dynamics of a chemoattractant (concentration $c$), which has diffusivity $D_{\text{c}}$ and is absorbed by each cell at a maximal rate $k$, and cells (concentration $\rho$), which bias their motion in response to a sensed chemoattractant gradient (\cref{Fig schematic}):
\begin{equation} \label{eq chemo}
\partial_t c = D_{\text{c}} \nabla^2 c - k \rho g(c).
\end{equation}
\begin{equation} \label{eq cells}
\partial_t \rho = -\bm{\nabla}\cdot \bm{J};\qquad \bm{J} = -D_\rho \bm{\nabla} \rho +  \rho \chi \bm{\nabla} f(c).
\end{equation}
Here, $g(c)$ describes how chemoattractant uptake is limited by its availability, modeled using Michaelis-Menten kinetics as $g(c) = c/(c+c_{\text{M}})$ with half-maximum concentration $c_{\text{M}}$. The cell concentration evolves through the flux $\bm{J}$, which has a diffusive contribution arising from undirected motion with an effective diffusivity $D_\rho$, and a chemotactic contribution arising from directed motion up the chemical gradient with a drift velocity $\bm{v}_{\text{c}} = \chi \bm{\nabla} f(c)$. The function $f(c)$ characterizes the ability of cells to sense the chemoattractant. For illustration purposes, we use the established logarithmic sensing function\cite{Kalinin2009,Fu2018,Cremer2019} $f(c) = \ln \left(\frac{1+c/c_-}{1+c/c_+}\right)$, with lower and upper characteristic concentrations $c_-$ and $c_+$ (\cref{Fig sensing-response}, red). The chemotactic coefficient $\chi$ describes the ability of the cells to migrate up the sensed chemoattractant gradient. In what follows, we determine front stability in terms of $f'(c)>0$ and $f''(c)<0$, regardless of the specific form of $f(c)$. Hence, our results can be generalized to other active systems employing different forms of sensing that also typically increase and eventually saturate with increasing stimulus.

While additional details (e.g., other chemicals, cellular proliferation) can also be introduced, here we focus on the minimal model of chemotactic fronts. Indeed, in excellent agreement with experiments, \cref{eq chemo,eq cells} give rise to a propagating pulse of cells \cite{Keller1971a,Fu2018,Seyrich2019,Cremer2019,Bhattacharjee2021a}. However, the full \cref{eq chemo,eq cells} cannot be solved analytically, precluding a generic analysis of front stability.

\textbf{Flat front.} To overcome this issue, we follow earlier work \cite{Wang2017a} and consider a simplified description of the pulse as a step profile with cell concentration $\rho_{\text{p}}$ moving along the $\hat{\bm{x}}$ axis at speed $v_0$: $\rho_0(s) = \rho_{\text{p}}\theta(-s)$ (\cref{Fig gradient-mechanism}, green). Here, $s \equiv x - v_0 t$ is the comoving coordinate, and $\theta$ is the Heaviside step function. We discuss the validity of this approximation in the \hyperref[SI]{SI}. The front of the pulse, located at $s=0$, is taken to be flat, i.e. independent of the transverse coordinate $y$ (dashed line in \cref{Fig schematic}). Ahead of the pulse ($s>0$), there are no cells, and hence no chemoattractant absorption. Inside the pulse ($s<0$), chemoattractant is absorbed; we assume that its concentration is smaller or similar to $c_{\text{M}}$, and hence we approximate $g(c)\approx c/c_{\text{M}}$, whose validity we verify \emph{a posteriori} using our parameter estimates (\cref{t parameters}). We impose boundary conditions $c(s\rightarrow -\infty) = 0$ and $c(s\rightarrow \infty) = c_\infty$, with $c_\infty$ being the chemoattractant concentration far ahead of the front, and we require continuity of the chemoattractant concentration and flux at the front. 

We thereby obtain the traveling chemoattractant profile $c_0(s)$ (\cref{Fig gradient-mechanism}, orange):
\begin{subequations} \label{eq chemo-flat-solutions}
\begin{align}
c_0^{\text{a}}(s) &= c_\infty \left[1 - \frac{\sqrt{1+ 4\Gamma} - 1}{\sqrt{1+4\Gamma} + 1} \exp\left[-\frac{s}{\ell_{\text{d}}}\right]\right];\quad s\ge 0, \label{eq ahead-solution}\\
c_0^{\text{i}}(s) &= \frac{2c_\infty}{\sqrt{1+4\Gamma} + 1}\exp\left[\frac{s}{2\ell_{\text{d}}} \left(\sqrt{1+ 4\Gamma} - 1\right)\right]; \quad s\le 0. \label{eq inside-solution}
\end{align}
\end{subequations}
Ahead of the pulse (\cref{eq ahead-solution}), the chemoattractant concentration varies exponentially over a diffusion length scale $\ell_{\text{d}}\equiv D_{\text{c}}/v_0$, which results from the balance of front motion and chemoattractant diffusion. Inside the pulse (\cref{eq inside-solution}), chemoattractant decays over a different, internal length scale $\ell_{\text{i}} \equiv \sqrt{\ell_{\text{d}} \ell_{\text{a}}} \equiv \ell_{\text{d}} /\sqrt{\Gamma}$, where the absorption length $\ell_{\text{a}} \equiv v_0 c_{\text{M}}/(k \rho_{\text{p}})$ results from the balance of front motion and chemoattractant absorption. We have also defined the dimensionless parameter $\Gamma\equiv \ell_{\text{d}}/\ell_{\text{a}}$, which we call the dif\-fu\-sio-ab\-sorp\-tion number. Representative values of all these parameters are given in \cref{t parameters}.

\textbf{Front perturbations.} We next analyze the linear stability of this front against morphological perturbations. We perturb the cell concentration profile along the $\bm{\hat{y}}$ axis, transverse to the propagation direction: $\rho(x,y,t) = \rho_0(s - \delta x_{\text{f}}(y,t))$, where $\delta x_{\text{f}}(y,t)$ represents the perturbation in front position (\cref{Fig schematic}). Consequently, the chemoattractant field is perturbed as $c(x,y,t) = c_0(s) + \delta c(s,y,t)$. For perturbations of wave number $q$, the chemoattractant field relaxes at a rate $\sim D_{\text{c}} q^2$ according to \cref{eq chemo}. We assume $D_{\text{c}} \gg D_\rho$, as is the case for cells migrating in porous media or on substrates. In this limit, chemoattractant perturbations rapidly reach a quasi-stationary profile $\delta c(s,y)$ that adapts to the slowly-evolving cell front (\hyperref[SI]{SI}). 

The cell front moves by diffusion and chemotaxis (\cref{eq cells}). As expected, the diffusive flux $-D_\rho \bm{\nabla}\rho$ tends to stabilize the front by smoothing out transverse gradients of cell concentration. The influence of the chemotactic drift flux $\rho \bm{v}_{\text{c}}$, however, is more subtle. To gain intuition, we express the chemotactic velocity as $\bm{v}_{\text{c}} = \chi \bm{\nabla} f(c) = \chi f'(c) \bm{\nabla}c$. As in linear response theory, $\bm{v}_{\text{c}}$ can be viewed as the cellular response to the driving force given by the chemoattractant gradient, $\bm{\nabla}c$, with $\chi f'(c)$ being the response function. Whereas the sensing ability $f(c)$ increases with chemoattractant concentration, the tactic response $f'(c)$ decreases as sensing becomes increasingly saturated (\cref{Fig sensing-response}). Because $\bm{v}_{\text{c}}$ involves the product of $f'(c)$ and $\bm{\nabla}c$, its perturbation has two contributions, $\delta \bm{v}_{\text{c}} = \chi \left[f'(c) \grad \delta c + \delta f'(c) \grad c \right]$, which correspond to perturbations of the gradient and the response, respectively.

\textbf{Competing mechanisms of front stability.} How do these distinct contributions affect front stability? In a linear stability analysis, to first order in perturbations, front motion depends on the chemotactic velocity perturbation $\delta \bm{v}_{\text{c}}$ evaluated at the position of the unperturbed front, $s=0$. While this perturbation has components both in the transverse ($\hat{\bm{y}}$) and the propagation ($\hat{\bm{x}}$) directions, as we show in the full analysis in the \hyperref[SI]{SI}, front stability is determined by the sign of the $\hat{\bm{x}}$ component,
\begin{equation} \label{eq drift-perturbations-front}
\delta v_{\text{c},x}(s=0,y) = \chi \left[ f'_0 \,\partial_s \delta c(0,y) + \partial_s c_0(0)\, f''_0 \,\delta c(0,y) \right].
\end{equation}
Here, we have used $\delta f'(c) = f''(c) \delta c$ and expressed dependencies on $x$ \textit{via} the comoving coordinate $s = x-v_0 t$; $c_0(s)$ is given by \cref{eq chemo-flat-solutions}.

The first contribution in \cref{eq drift-perturbations-front} is given by changes in the chemoattractant gradient at the position of the unperturbed front, $\partial_s \delta c(0,y)$, multiplied by the unperturbed chemotactic response, $f'_0 \equiv f'(c_0(0))>0$. We name this contribution the \textit{gradient mechanism}; it represents changes in cell velocity due to spatial variations in the driving force $\bm{\nabla}c$. Specifically, in peaks of the perturbed front ($\delta x_{\text{f}}(y)>0$), cells populate the position of the unperturbed front ($s=0$), thereby absorbing chemoattractant and decreasing its concentration: $\delta c(0,y)<0$ (compare peak and flat in \cref{Fig gradient-mechanism}). As a result, the chemoattractant gradient inside the pulse ($s<0$) decreases with respect to the unperturbed situation (\cref{Fig gradient-mechanism}), and thus $\partial_s \delta c(0,y)<0$. Because this first contribution in \cref{eq drift-perturbations-front} is negative, it is stabilizing. Intuitively, the decrease in chemoattractant gradient slows down cells in peaks, allowing the rest of the population to catch up and flatten the front.

The second contribution in \cref{eq drift-perturbations-front} is given by the unperturbed chemoattractant gradient, $\partial_s c_0(0)>0$, multiplied by the change in the chemotactic response at the front, $\delta f'(c) = f''_0 \delta c$, where $f''_0 \equiv f''(c_0(0))<0$ (\cref{Fig sensing-response}). We name this contribution the \textit{response mechanism}; it represents changes in cell velocity due to spatial variations in the cells' chemotactic response. As noted above, in peaks of the perturbed front ($\delta x_{\text{f}}(y)>0$), cells absorb chemoattractant and decrease its concentration at $s=0$, giving $\delta c(0,y)<0$. Because this second contribution in \cref{eq drift-perturbations-front} is positive, it is destabilizing. Intuitively, the decrease in chemoattractant causes cells in peaks to respond to the gradient more strongly (compare peak and flat in \cref{Fig response-mechanism}) and move faster, leaving the rest of the population behind and amplifying front perturbations.

Thus, our analysis reveals two competing chemotactic mechanisms that determine front stability: Cells at a bulging part of the front are exposed to a smaller chemoattractant gradient, which slows them down (gradient mechanism), but they respond more strongly to the gradient, which speeds them up (response mechanism). To quantitatively compare these two mechanisms, we rewrite \cref{eq drift-perturbations-front} as $\delta v_{\text{c},x}(s=0,y) = \chi \left[ \alpha \,\partial_s  - \beta \frac{c'_0(0)}{c_\infty} \right]\frac{\delta c(0,y)}{c_\infty}$, where the two positive dimensionless parameters $\alpha \equiv f'_0 c_\infty$ and $\beta \equiv - f''_0 c_\infty^2$ quantify the strengths of the gradient and response mechanisms, respectively.

\begin{figure}[tb]
\begin{center}
\includegraphics[width=\columnwidth]{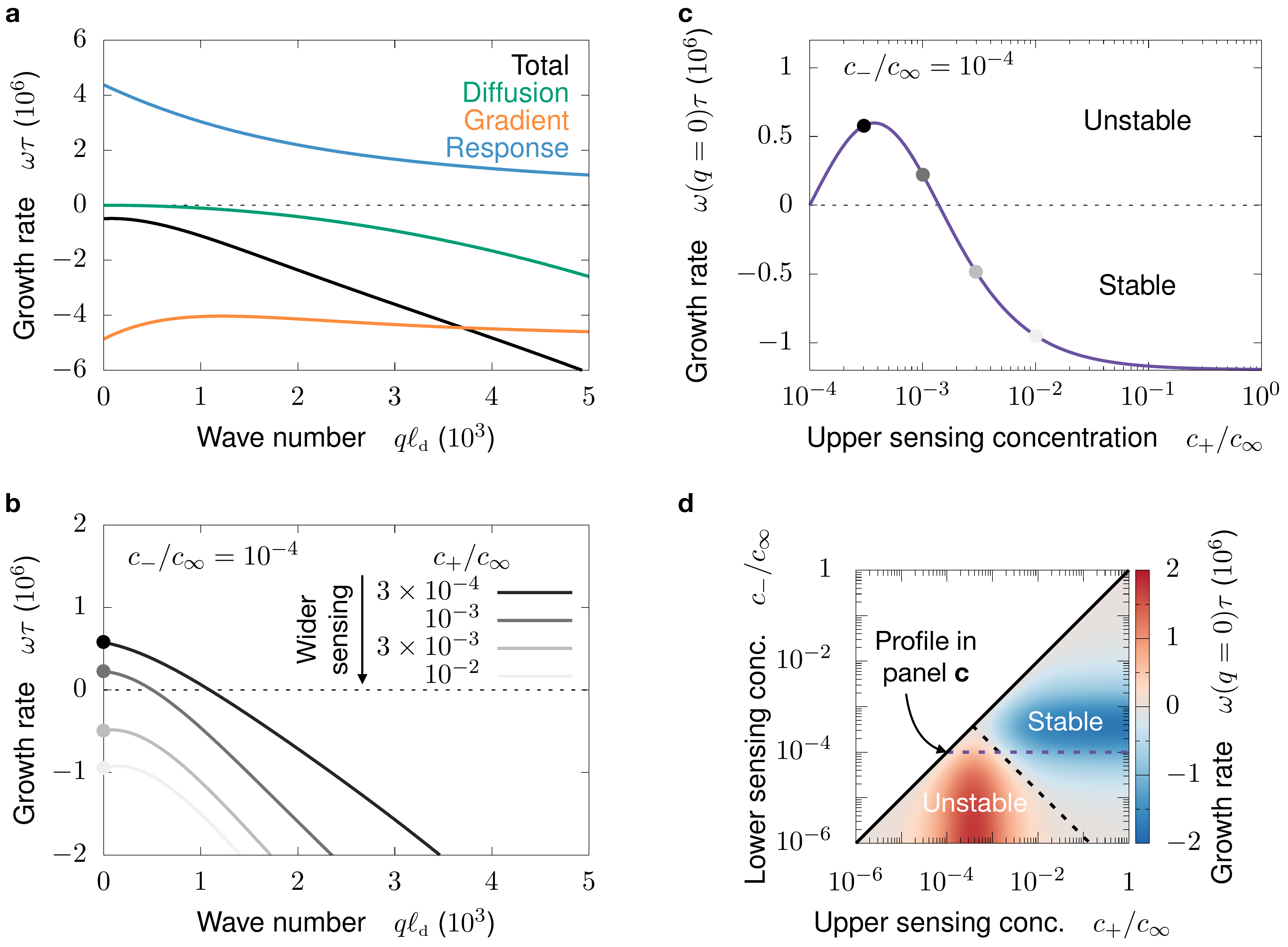}
\end{center}
  {\phantomsubcaption\label{Fig growth-rate}}
  {\phantomsubcaption\label{Fig sensing-window}}
  {\phantomsubcaption\label{Fig growth-rate-upper}}
  {\phantomsubcaption\label{Fig stability-diagram}}
\bfcaption{Cellular sensing governs chemotactic front stability}{ \subref*{Fig growth-rate}, Growth rate of front perturbations, showing the contributions of cell diffusion as well as the gradient and response chemotactic mechanisms (see \cref{Fig gradient-mechanism,Fig response-mechanism}). \subref*{Fig sensing-window}, Increasing the upper sensing concentration $c_+$ promotes front stability. \subref*{Fig growth-rate-upper}, As sensing becomes less limited at higher concentrations by increasing $c_+$, the front can switch from unstable to stable, as indicated by the sign of the long-wavelength growth rate $\omega(q=0)$. The points correspond to those in panel \subref*{Fig sensing-window}. \subref*{Fig stability-diagram}, Diagram of front stability as a function of the lower and upper characteristic sensing concentrations. The color code informs about the degree of front stability, as given by $\omega(q=0)$. The black dashed line indicates the stability limit (\cref{eq stability-sensing}). The purple dashed line indicates the slice of the diagram shown in panel \subref*{Fig growth-rate-upper}. Throughout the figure, the growth rate is rescaled by what we call the chemotactic time $\tau \equiv \ell_{\text{d}}^2/\chi$. Parameter values are in \cref{t parameters}. } \label{Fig3}
\end{figure}

\textbf{Chemotactic fingering instability.} Having identified the two mechanisms whereby chemotaxis influences front stability, we solve the full \cref{eq cells} to obtain front speed perturbations $\delta v(y,t)=\partial_t \delta x_{\text{f}}(y,t)$ (\cref{eq velocity-perturbations-Fourier}), and hence the growth rate $\omega(q) \equiv \delta\tilde v(q)/\delta\tilde x_{\text{f}}(q)$ of front perturbations with wave number $q$, where tildes indicate Fourier components (\hyperref[SI]{SI}):
\begin{multline} \label{eq growth-rate}
\omega(q) = -D_\rho q^2 + \frac{\chi}{\ell_{\text{d}}^2} \frac{\sqrt{1+4\Gamma} - 1}{\sqrt{1 + 4q^2 \ell_{\text{d}}^2} + \sqrt{1 + 4(\Gamma + q^2 \ell_{\text{d}}^2)}}\\
\times \left[\beta \frac{\sqrt{1+ 4\Gamma} - 1}{\sqrt{1+ 4\Gamma} + 1} - \frac{\alpha}{2} \left(\sqrt{1+4(\Gamma+q^2\ell_{\text{d}}^2)} - 1\right) \phantom{\frac{q^2}{\sqrt{q^2}}}\right.\\
\left. - 2\alpha \frac{q^2 \ell_{\text{d}}^2}{\sqrt{1+4(\Gamma+q^2\ell_{\text{d}}^2)} - 1}\right].
\end{multline}
\cref{Fig growth-rate} shows this growth rate separating the contributions of the different mechanisms. As expected, the diffusive contribution $-D_\rho q^2$ is always stabilizing (\cref{Fig growth-rate}, green). At large length scales (small $q$), it is negligible in front of the two chemotactic mechanisms resulting from \cref{eq drift-perturbations-front}. In agreement with our argument above, the \textit{gradient mechanism} ($\propto \alpha$ in \cref{eq growth-rate}) is stabilizing (\cref{Fig growth-rate}, orange), while the \textit{response mechanism} ($\propto \beta$ in \cref{eq growth-rate}) is destabilizing (\cref{Fig growth-rate}, blue). In the long-wavelength limit ($q\rightarrow 0$), we have $\omega(0) = \frac{\chi}{\ell_{\text{d}}^2} \left(\frac{\sqrt{1+4\Gamma} - 1}{\sqrt{1+4\Gamma} + 1}\right)^2 \left[ \beta - \frac{\alpha}{2} \left(\sqrt{1+ 4\Gamma} + 1\right)\right]$, and hence the flat front becomes unstable, $\omega(0)>0$, if
\begin{equation} \label{eq stability}
\beta > \frac{\alpha}{2} \left(\sqrt{1+ 4\Gamma} + 1\right),
\end{equation}
i.e. if the chemotactic response decreases too strongly with chemoattractant concentration, corresponding to large values of $\beta$. In this case, cells at valleys, which are exposed to higher concentrations, respond too weakly and are left behind by cells at peaks, which are instead exposed to lower concentrations and thus respond more strongly to the gradient.

\textbf{Cellular sensing governs chemotactic front stability.} Our central result, given by \cref{eq growth-rate,eq stability}, is that the limited ability of single cells to sense high concentrations of chemoattractant, and the resulting limitation in their chemotactic response, can destabilize entire propagating fronts. To illustrate this, we recast our results in terms of the characteristic concentrations $c_-$ and $c_+$ of the sensing function $f(c)$. Varying these concentrations tunes both $f'(c)$ and $f''(c)$, thus affecting the values of both $\alpha \equiv f'_0 c_\infty$ and $\beta \equiv - f''_0 c_\infty^2$, and hence changing the relative contribution of the stabilizing and the destabilizing mechanisms. Which effect wins when varying $c_-$ and $c_+$?

For a given $c_-$, the front is unstable for values of $c_+$ close to $c_-$, i.e. for narrow sensing windows (darker curves in \cref{Fig sensing-window}). As $c_+$ increases, the destabilizing effect of the chemotactic response limitation becomes less important, and the front eventually becomes stable (lighter curves in \cref{Fig sensing-window}). Therefore, for a given $c_-$, the front switches from unstable to stable as the sensing window widens by increasing $c_+$ (\cref{Fig growth-rate-upper}, corresponding to the purple dashed line in \cref{Fig stability-diagram}). Conversely, the front can also be stabilized by narrowing the sensing window, e.g. by increasing $c_-$ at fixed $c_+$ (moving up in \cref{Fig stability-diagram}). Therefore, front stability is promoted by increasing the characteristic sensing concentrations $c_-$ and $c_+$. Although increasing $c_-$ and $c_+$ weakens the chemotactic response (\cref{Fig sensing-response}), it also makes the destabilizing response-limitation effects less pronounced. Finally, we recast the instability condition, \cref{eq stability}, in terms of $c_-$ and $c_+$:
\begin{equation} \label{eq stability-sensing}
\frac{c_+}{c_\infty} < \frac{c_\infty}{c_-} \frac{2}{1 + 2\Gamma + \sqrt{1 + 4\Gamma}}.
\end{equation}
The black dashed line in \cref{Fig stability-diagram} shows the stability limit.

To test our predictions, we perform finite-element simulations of the full \cref{eq chemo,eq cells} (\hyperref[SI]{SI}). Introducing an initial perturbation with a long wavelength ($q\ell_{\text{d}} = 0.02$) and small amplitude ($A/\lambda = 0.016$), we find regimes of both front instability and stability (\cref{Fig examples}) in agreement with the stability diagram predicted analytically (\cref{Fig simulations-diagram}).

\textbf{Discussion.} We have quantified the conditions for the stability of chemotactic fronts. Below the stability limit (\cref{eq stability-sensing}), we predict a morphological instability that could result in fingering patterns and even front disassembly. Expanding bacterial colonies form complex patterns that are thought to arise from bulk instabilities \cite{Murray2003,Budrene1991,Budrene1995,Brenner1998,Cates2010,Brenner2010}. However, the instability that we predict is fundamentally different, as it is interfacial and it arises purely from chemotaxis. To our knowledge, it has not been observed in experiments. Our predictions provide guidelines for future studies to search for it. For example, we predict front instability when the sensing concentrations $c_-$ and $c_+$ are small compared to nutrient availability $c_\infty$ (\cref{Fig stability-diagram}). Therefore, experiments can probe this regime by either increasing nutrient availability or genetically impairing the cells' sensing ability. Moreover, we predict that fronts would destabilize over long wavelengths, at least of the order of the diffusion length $\ell_{\text{d}}$ (\cref{Fig sensing-window}). In experiments, fronts must therefore be sufficiently long to become unstable.

Front stability can have relevant biological implications. For example, in embryos, chemotactic cell groups must remain cohesive to develop into functional organs. In bacterial populations, cells must also stay together to collectively absorb sufficient chemoattractant to generate the chemical gradient driving front motion \cite{Tweedy2016,Tweedy2020a}. Thus, inspired by our calculations, we speculate that the cells' sensing abilities might have evolved to avoid instability and ensure robust collective chemotaxis.

To probe this idea, we examine published experiments on chemotactic fronts of \textit{E. coli} \cite{Fu2018,Bhattacharjee2021a}. These experiments report the concentrations $c_-$ and $c_+$ for two different chemoattractants, as well as the parameters that determine the diffusio-absorption number $\Gamma$ (\cref{t sensing-estimates}), with which we construct a stability diagram akin to \cref{Fig stability-diagram} for each experiment (\cref{Fig experiment}). The far-field concentrations $c_\infty$ used in experiments likely represent upper bounds of those encountered in natural environments, and thus our estimates are in conditions most favorable for \emph{instability}. Yet, we find that all experiments fall in the predicted \emph{stable} regime. Consistently, the experiments observe stable flat fronts in all cases, suggesting that the ratios $c_-/c_\infty$ and $c_+/c_\infty$ are always high enough (\cref{Fig experiment}). Further experiments are required to systematically test the tantalizing hypothesis that cellular sensing might be tuned to ensure stable collective chemotaxis. 

Our results are also qualitatively consistent with recent experiments on 3D-printed bacterial populations, which found that morphological perturbations are smoothed out by chemotaxis \cite{Bhattacharjee2022}. These experiments, however, imposed large-amplitude perturbations in three-dimensional populations, whereas our analysis focuses on the small-amplitude limit in two dimensions. Hence, the experiments cannot be directly compared to our theory. Nevertheless, both demonstrate that sensing limitations of individual cells determine the stability of an entire chemotactic population. 

Building on this finding, future work can explore how population morphology is affected by the chemotactic efficiency constraints imposed by biochemical \cite{Camley2018,Ellison2016,Mugler2016,Fancher2017} and mechanical \cite{Colin2019,Tian2021} cell-cell interactions, switching between swimming states \cite{Alirezaeizanjani2020}, and information acquisition requirements \cite{Mattingly2021}. Our work could also be generalized to account for collective sensing mechanisms \cite{Camley2018} and for chemokinesis, i.e. the dependence of cell speed on chemical concentration \cite{Jakuszeit2021}. Furthermore, whereas the instability mechanisms that we identified arise from the deterministic dynamics of chemotaxis, future work can study the role of noise in selecting the resulting patterns. Beyond chemotaxis, our theory could be generalized to other types of collective tactic phenomena \cite{Roca-Cusachs2013,Shellard2020a,Sengupta2021} including cell durotaxis \cite{Sunyer2016,Alert2019a}, electrotaxis \cite{Cohen2014}, and robot phototaxis \cite{Mijalkov2016,Palagi2018}. In these cases, as for chemotaxis, sensing increases and then saturates with the stimulus, be it substrate stiffness \cite{Alert2019a,Ghibaudo2008}, electric field \cite{Cohen2014}, or light intensity \cite{Mijalkov2016,Palagi2018} --- which, as quantified by the sensing function $f(c)$, is the essential feature of our theory. Specifically, in our analysis of chemotactic front propagation in terms of linear response theory, chemical gradients provide the driving force, and cellular sensing provides the response function. In these general terms, we conclude that, when modulated by a response function, the force that drives front propagation can also fully determine its stability.\\

\textbf{Acknowledgments.} We thank D.B. Amchin, T. Bhattacharjee, and N.S. Wingreen for useful discussions. R.A. acknowledges support from the Princeton Center for Theoretical Science, from the Human Frontier Science Program (LT000475/2018-C), and from the National Science Foundation through the Center for the Physics of Biological Function (PHY-1734030). A.M.-C. acknowledges support from the Princeton Center for Theoretical Science and the Human Frontier Science Program through the grant LT000035/2021-C. S.S.D. acknowledges support from NSF grant CBET-1941716, the Eric and Wendy Schmidt Transformative Technology Fund at Princeton, and the Princeton Center for Complex Materials, a Materials Research Science and Engineering Center supported by NSF grant DMR-2011750.

\bibliography{Fronts}

\onecolumngrid %this command here balances the columns before moving to the next page

\clearpage
%\appendix

\setcounter{equation}{0}
\setcounter{figure}{0}
\renewcommand{\theequation}{S\arabic{equation}}
\renewcommand{\thefigure}{S\arabic{figure}}

\onecolumngrid
\begin{center}
\textbf{\large Supplementary Material for ``Cellular Sensing Governs the Stability of Chemotactic Fronts''}\\

\bigskip

Ricard Alert, Alejandro Mart\'{i}nez-Calvo, and Sujit S. Datta

\bigskip

\end{center}

\twocolumngrid

\label{SI}

\section{The step-profile approximation} \label{step-profile}

The step-profile approximation (\cref{Fig gradient-mechanism}, green) is based on two assumptions:
\begin{enumerate}[(a)]
\item the cell concentration remains uniform inside the pulse ($s<0$), and
\item the cell concentration drops suddenly to zero ahead of the pulse ($s>0$).
\end{enumerate}

Assumption (a) is often fulfilled in experiments. For example, for the experiments of Ref.\cite{Bhattacharjee2021a} (parameter values in \cref{t parameters}), the cell pulse is about $700$ $\mu$m wide whereas the chemoattractant penetrates only about $\ell_{\text{i}}\approx 7$ $\mu$m into the pulse. The pulse can therefore be approximated as infinitely wide, with an approximately uniform cell concentration across the region where the chemoattractant concentration varies. More generally, from the classic work of Keller and Segel in Ref.\cite{Keller1971}, the pulse width is of the order of $w\sim D_\rho/v_0$. Assumption (a) then holds as long as $w/\ell_{\text{i}}\gg 1$. Using the definition of $\ell_{\text{i}}$ as listed in \cref{t parameters}, this condition amounts to $D_\rho/v_0 \times  \sqrt{k \rho_{\text{f}}/(D_{\text{c}}  c_{\text{M}})}\gg 1$. Therefore, assumption \textit{a} is a controlled approximation.

In contrast, assumption (b) is uncontrolled. We thus test its validity in numerical solutions of the full \cref{eq chemo,eq cells}, as described in \cref{numerical}. We find that, from the initial condition, the decay of cell concentration ahead of the front becomes even sharper when the front is transversely unstable, and it smoothes out when the front is stable \cref{Fig profiles}. Therefore, the step-profile approximation is particularly applicable to conditions of front instability.

Overall, the step-profile approximation allows us to make analytical progress and thereby identify the physical mechanisms of front stability. The competing mechanisms that we identify, i.e. the gradient and the response mechanisms, are general conclusions that hold irrespective of the accuracy of our calculation.

\section{Linear stability analysis} \label{LSA}

Here, we provide the details of the linear stability analysis of chemotactic fronts. As explained in the Main Text, we impose a perturbation $\delta x_{\text{f}}(y,t)$ on the front position, corresponding to a traveling cell density field $\rho(x,y,t) = \rho_0(s - \delta x_{\text{f}}(y,t))$, which has the same step profile as for the unperturbed front. As a result of the front perturbations, the chemoattractant field will be $c(x,y,t) = c_0(s) + \delta c(s,y,t)$, whose perturbation will rapidly reach a quasi-stationary traveling profile $\delta c(s,y)$ that follows the slowly-evolving cell front.

\textbf{Chemoattractant perturbations.} As for the unperturbed chemoattractant profile in \cref{eq chemo-flat-solutions}, we determine $\delta c(s,y)$ separately in the regions ahead and inside the cell pulse. Ahead of the pulse, there is no chemoattractant absorption, and hence \cref{eq chemo} reduces to
\begin{equation} \label{eq ahead}
\partial_t c^{\text{a}} = D_{\text{c}} \nabla^2 c^{\text{a}}.
\end{equation}
Respectively, inside the pulse, cell concentration is uniform, $\rho = \rho_{\text{p}}$. Approximating the chemoattractant uptake function by $g(c) \approx c/c_{\text{M}}$ as explained in the Main Text, \cref{eq chemo} inside the pulse reduces to
\begin{equation} \label{eq inside}
\partial_t c^{\text{i}} = D_{\text{c}} \nabla^2 c^{\text{i}} - \frac{k \rho_{\text{p}}}{c_{\text{M}}} c^{\text{i}},
\end{equation}
Assuming a propagating solution $c(x,y,t) = c(x-v_0 t,y)$, linearizing \cref{eq ahead,eq inside}, and expressing them in terms of the comoving coordinate $s = x-v_0 t$, we obtain
\begin{subequations} \label{eq chemo-perturbations}
\begin{align}
- v_0\, \partial_s \delta c^{\text{a}}(s,y) &= D_{\text{c}} (\partial_s^2 + \partial_y^2) \delta c^{\text{a}}(s,y), \\
- v_0\, \partial_s \delta c^{\text{i}}(s,y) &= D_{\text{c}} (\partial_s^2 + \partial_y^2) \delta c^{\text{i}}(s,y) - \frac{k \rho_{\text{p}}}{c_{\text{M}}} \delta c^{\text{i}}(s,y).
\end{align}
\end{subequations}

To solve these equations, we decompose $\delta c(s,y)$ in Fourier modes along the $\hat{\bm{y}}$ axis:
\begin{equation}
\delta c(s,y) = \int_{-\infty}^\infty \delta\tilde{c} (s,q) \,e^{iqy} \frac{\dd q}{2\pi}.
\end{equation}
In Fourier components, \cref{eq chemo-perturbations} becomes
\begin{subequations}
\begin{align}
- v_0\, \partial_s \delta\tilde c^{\text{a}} & = D_{\text{c}} (\partial_s^2 - q^2) \delta\tilde c^{\text{a}},\\
- v_0\, \partial_s \delta\tilde c^{\text{i}} & = D_{\text{c}} (\partial_s^2 - q^2) \delta\tilde c^{\text{i}} - \frac{k \rho_{\text{p}}}{c_{\text{M}}} \delta\tilde c^{\text{i}}.
\end{align}
\end{subequations}
The propagating solutions to these equations are
\begin{subequations} \label{eq chemo-perturbations-results}
\begin{align}
\label{eq chemo-perturbation-ahead}
\delta\tilde c^{\text{a}}(s,q) &= C_{\text{a}} \exp\left[ -\frac{s}{2\ell_{\text{d}}} \left( 1 + \sqrt{1 + 4 q^2 \ell_{\text{d}}^2}\right)\right], \\
\delta\tilde c^{\text{i}}(s,q) &= C_{\text{i}} \exp\left[ \frac{s}{2\ell_{\text{d}}} \left( \sqrt{1 + 4(\Gamma + q^2 \ell_{\text{d}}^2)} - 1\right)\right]. \label{eq chemo-perturbation-inside}
\end{align}
\end{subequations}
Here, as explained in the Main Text, we have defined $\ell_{\text{d}} \equiv D_{\text{c}}/v_0$ and $\Gamma\equiv \ell_{\text{d}}/\ell_{\text{a}}$, with $\ell_{\text{a}} \equiv v_0 c_{\text{M}}/(k\rho_{\text{p}})$. Moreover, we have already set two integration constants by imposing $\delta c^{\text{a}}(s\rightarrow \infty,y) = 0$ and $\delta c^{\text{i}}(s\rightarrow -\infty,y)= 0$. The remaining two integration constants, $C_{\text{a}}$ and $C_{\text{i}}$, are determined by imposing equality of both chemoattractant concentration and diffusive flux at the perturbed front: $c^{\text{a}}(\delta x_{\text{f}}(y),y) = c^{\text{i}}(\delta x_{\text{f}}(y,t),y)$ and $\partial_s c^{\text{a}}(\delta x_{\text{f}}(y),y) = \partial_s c^{\text{i}}(\delta x_{\text{f}}(y,t),y)$. Expanding these conditions to first order in perturbations, we obtain
\begin{subequations}
\begin{align}
\delta c^{\text{a}}(0) &= \delta c^{\text{i}}(0),\\
\partial_s \delta c^{\text{a}}(0) + \partial_s^2 c_0^{\text{a}}(0)\, \delta x_{\text{f}} &= \partial_s \delta c^{\text{i}}(0) + \partial_s^2 c_0^{\text{i}}(0)\, \delta x_{\text{f}},
\end{align}
\end{subequations}
which give
\begin{equation} \label{eq integration-constants}
C_{\text{a}} = C_{\text{i}} = -c_\infty \frac{\sqrt{1 + 4\Gamma} - 1}{\sqrt{1 + 4q^2 \ell_{\text{d}}^2} + \sqrt{1 + 4(\Gamma + q^2 \ell_{\text{d}}^2)}} \frac{\delta\tilde{x}_{\text{f}}}{\ell_{\text{d}}}.
\end{equation}
For $\Gamma\gg 1$ as in our parameter estimates (\cref{t parameters}), the chemoattractant perturbation profile can be approximated as
\begin{subequations} \label{eq chemo-perturbations-largeA}
\begin{align}
%\begin{gather}
\label{eq chemo-perturbation-ahead-largeA}
&\delta\tilde c^{\text{a}}(s,q) \approx C_{\text{a}} \exp\left[ -\frac{s}{2\ell_{\text{d}}} \left( 1 + \sqrt{1 + 4 q^2 \ell_{\text{d}}^2}\right)\right],\\
&\delta\tilde c^{\text{i}}(s,q) \approx C_{\text{i}} \exp\left[ \frac{s}{\ell_{\text{d}}} \sqrt{\Gamma + q^2 \ell_{\text{d}}^2}\right],
%\end{gather}
\end{align}
\end{subequations}
with
\begin{equation} \label{eq integration-constants-largeA}
C_{\text{a}} = C_{\text{i}} \approx -c_\infty \frac{2\sqrt{\Gamma}}{\sqrt{1 + 4q^2 \ell_{\text{d}}^2} + 2\sqrt{\Gamma + q^2 \ell_{\text{d}}^2}} \frac{\delta\tilde x_{\text{f}}}{\ell_{\text{d}}}.
\end{equation}

\textbf{Interpretation.} To interpret these results, let's consider a sinusoidal front perturbation of amplitude $\delta A$ and wavenumber $k$: $\delta x_{\text{f}}(y) = \delta A \sin(ky)$. For this perturbation, we obtain $\delta\tilde{x}_{\text{f}}(q)$, introduce it in \cref{eq chemo-perturbations-results} via \cref{eq integration-constants}, and perform an inverse Fourier transform to obtain the chemoattractant perturbations in real space:
\begin{subequations} \label{eq chemo-perturbations-real}
\begin{gather}
\begin{multlined}
\label{eq chemo-perturbation-real-ahead}
\delta c^{\text{a}}(s,y) = - C \exp\left[ -\frac{s}{2\ell_{\text{d}}} \left( 1 + \sqrt{1 + 4 k^2 \ell_{\text{d}}^2}\right)\right] \\
\times \frac{\delta A}{\ell_{\text{d}}} \sin(ky),
\end{multlined} \\
\begin{multlined}
\delta c^{\text{i}}(s,y) = - C \exp\left[\frac{s}{2\ell_{\text{d}}} \left(\sqrt{1 + 4(\Gamma + k^2 \ell_{\text{d}}^2)} - 1\right)\right] \\
\times \frac{\delta A}{\ell_{\text{d}}} \sin(ky), \label{eq chemo-perturbation-real-inside}
\end{multlined}
\end{gather}
\end{subequations}
where
\begin{equation}
    C = c_\infty \frac{\sqrt{1 + 4\Gamma} - 1}{\sqrt{1 + 4k^2 \ell_{\text{d}}^2} + \sqrt{1 + 4(\Gamma + k^2 \ell_{\text{d}}^2)}}
\end{equation}
is a positive constant. These results show that in regions around peaks (\cref{Fig schematic}), where the front protrudes outward ($\sin(ky)>0$), chemoattractant becomes absorbed, and its concentration decreases ($\delta c<0$). Conversely, in regions around valleys (\cref{Fig schematic}), where the front bends inward ($\sin(ky)<0$), chemoattractant becomes replenished and its concentration increases ($\delta c>0$). Hence, in protruding regions, the chemoattractant gradient is increased ahead of the front ($s>0$) but decreased inside the front ($s<0$); compare peak and flat in \cref{Fig gradient-mechanism}. Respectively, in intruding regions, the chemoattractant gradient is decreased ahead of the front ($s>0$) but increased inside the front ($s<0$); compare valley and flat in \cref{Fig gradient-mechanism}.

\textbf{Growth rate of front perturbations.} Front perturbations evolve at a rate given by the perturbation in front speed $\delta v(y,t) = \partial_t \delta x_{\text{f}}(y,t)$. To obtain $\delta v$, we linearize the cell concentration dynamics \cref{eq cells}. For the imposed traveling solution $\rho(x,y,t) = \rho_0(s-\delta x_{\text{f}}(y,t))$, and in terms of the comoving coordinate $s = x-v_0 t$, we obtain:
\begin{multline} \label{eq velocity-perturbations}
-\delta v\, \partial_s \rho_0 = -\partial_s \left(\rho_0 \chi \left[ f'(c_0) \,\partial_s \delta c + f''(c_0)\, \partial_s c_0\, \delta c\right]\right)\\
 - \rho_0 \chi f'(c_0)\, \partial_y^2 \delta c - D_\rho\, \partial_s \rho_0\, \partial_y^2 \delta x_{\text{f}}.
\end{multline}
The left-hand-side term is the advective flux due to perturbations in front motion. The right-hand-side term on the first line, with the $\partial_s$ derivative, results from chemotactic fluxes in the propagation direction ($\hat{\bm{x}}$). Respectively, the terms on the second line correspond to fluxes in the transverse direction ($\hat{\bm{y}}$), which stem from both directed (chemotactic, $\chi$) and undirected (diffusive, $D_\rho$) cell motion.

To solve for $\delta v$, we integrate \cref{eq velocity-perturbations} over $s$, taking into account that $\rho_0$ corresponds to a step profile, and keeping terms only to first order in perturbations. We then transform to Fourier space and obtain
\begin{multline} \label{eq velocity-perturbations-Fourier}
\delta\tilde v = \chi \left[f'_0\, \partial_s \delta\tilde c(0,q) + f''_0\, \partial_s c_0(0)\, \delta\tilde c(0,q) \phantom{- q^2 \int_{-\infty}^0 f'(c_0(s)) \delta \tilde c(s,q) \;\dd s} \right.\\
\left. - q^2 \int_{-\infty}^0 f'(c_0(s))\, \delta \tilde c(s,q) \;\dd s\right] - D_\rho q^2 \delta\tilde x_{\text{f}},
\end{multline}
where $f'_0 \equiv f'(c_0(0))>0$, $f''_0 \equiv f''(c_0(0))<0$ are the slope and curvature of the sensing function $f(c)$ at the unperturbed front ($s=0$). To complete the calculation of $\delta\tilde{v}$, we introduce our previous result for the chemoattractant perturbations $\delta\tilde{c}(s,q)$ (\cref{eq chemo-perturbations-results}). Note that the chemoattractant gradient perturbation must be evaluated inside the pulse, where there are cells, as opposed to cell-free region ahead of the front. Therefore, $\partial_s \delta\tilde{c}(0,q) = \lim_{s\rightarrow 0^-} \partial_s \delta\tilde{c}(s,q)$, which we evaluate using \cref{eq chemo-perturbation-inside}. For the same reason, the integral in \cref{eq velocity-perturbations-Fourier} runs only up to $s=0$.

Finally, to obtain the growth rate $\omega(q)$ of front perturbation modes, we use that $\delta v(y,t) = \partial_t \delta x_{\text{f}}(y,t)$. In Fourier space, we have $\delta\tilde v(q) = \omega(q) \delta\tilde x_{\text{f}}(q)$, and therefore the growth rate is $\omega(q) = \delta\tilde v(q)/\delta\tilde x_{\text{f}}(q)$. To obtain a closed analytical expression for the growth rate, we approximate the integral in \cref{eq velocity-perturbations-Fourier} by $f'_0 \int_{-\infty}^0 \delta\tilde{c}(s,q)\,\dd s$, which overestimates the contribution of the transverse chemotactic flux. Then, introducing the chemoattractant perturbations \cref{eq chemo-perturbations-results} with \cref{eq integration-constants}, we obtain
\begin{multline} \label{eq growth-rate-SI}
\omega(q) = -D_\rho q^2 + \frac{\chi}{\ell_{\text{d}}^2} \frac{\sqrt{1+4\Gamma} - 1}{\sqrt{1 + 4q^2 \ell_{\text{d}}^2} + \sqrt{1 + 4(\Gamma + q^2 \ell_{\text{d}}^2)}}\\
\times \left[\beta \frac{\sqrt{1+ 4\Gamma} - 1}{\sqrt{1+ 4\Gamma} + 1} - \frac{\alpha}{2} \left(\sqrt{1+4(\Gamma+q^2\ell_{\text{d}}^2)} - 1\right) \phantom{\frac{q^2}{\sqrt{q^2}}}\right.\\
\left. - 2\alpha \frac{q^2 \ell_{\text{d}}^2}{\sqrt{1+4(\Gamma+q^2\ell_{\text{d}}^2)} - 1}\right],
\end{multline}
which we quote in \cref{eq growth-rate} in the Main Text. Here, we have expressed $f'_0$ and $f''_0$ in terms of their corresponding positive dimensionless numbers $\alpha = f'_0 c_\infty$ and $\beta = -f''_0 c_\infty^2$, as explained in the Main Text. In the long-wavelength limit $q\rightarrow 0$, the growth rate tends to
\begin{equation} \label{eq zero-mode-SI}
\omega(0) = \frac{\chi}{\ell_{\text{d}}^2} \left(\frac{\sqrt{1+4\Gamma} - 1}{\sqrt{1+4\Gamma} + 1}\right)^2 \left[ \beta - \frac{\alpha}{2} \left(\sqrt{1+ 4\Gamma} + 1\right)\right],
\end{equation}
as we quote and discuss in the Main Text. Finally, in the limit $\Gamma\gg 1$ corresponding to our parameter estimates (\cref{t parameters}), these results are approximated as
\begin{multline} \label{eq growth-rate-approx}
\omega(q) \approx -D_\rho q^2 + \frac{\chi}{\ell_{\text{d}}^2} \frac{2\sqrt{\Gamma}}{\sqrt{1 + 4q^2 \ell_{\text{d}}^2} + 2\sqrt{\Gamma + q^2\ell_{\text{d}}^2}}\\
\times\left[\beta - \alpha \sqrt{\Gamma + q^2 \ell_{\text{d}}^2} - \alpha \frac{q^2\ell_{\text{d}}^2}{\sqrt{\Gamma + q^2\ell_{\text{d}}^2}} \right],
\end{multline}
and
\begin{equation} \label{eq zero-mode-approx}
\omega(0) \approx \frac{\chi}{\ell_{\text{d}}^2} \left[\beta - \alpha \sqrt{\Gamma}\right].
\end{equation}

\section{Numerical simulations} \label{numerical}

We perform two-dimensional (2D) numerical simulations of the full \cref{eq cells,eq chemo}, only with the approximation $g(c) \approx c/c_{\text{M}}$ as discussed in the Main Text. To this end, we make the equations dimensionless using the characteristic scales deduced in the Main Text, which define the following dimensionless variables for the position vector, time, and the cell density and chemoattractant concentration fields:
\begin{equation}
\tilde{\bm{r}} = \frac{\bm{r}}{\ell_{\dd}} = \frac{\bm{r} v_0}{D_{\textrm{c}}}, \quad \tilde{t} = \frac{t}{\tau} = \frac{t \chi}{\ell_{\dd}^2}, \quad \tilde{c} = \frac{c}{c_{\infty}}, \quad \tilde{\rho} = \frac{\rho}{\rho_{\textrm{p}}}.    
\end{equation}
In these dimensionless variables, \cref{eq chemo,eq cells} read
\begin{subequations}
\begin{gather}
\partial_{\tilde{t}} \tilde{c} = \frac{D_{\text{c}}}{\chi} \left[\nabla^2 \tilde{c} - \Gamma \tilde{\rho} \tilde{c}\right], \quad \text{in} \quad \V, \\
\partial_{\tilde{t}} \tilde{\rho} = \frac{D_{\rho}}{\chi} \nabla^2 \tilde{\rho} - \bm{\nabla} \cdot \left[ \tilde{\rho} \,\bm{\nabla} f(\tilde{c}) \right], \quad \text{in} \quad \V,\label{eq:bacterial}
\end{gather}
\end{subequations}
where $\V = (0,\tilde L_x)\times(0,\tilde L_y)$ is the 2D domain, $f(\tilde{c}) = \ln\left[\frac{1+\tilde{c}/(c_{-}/c_{\infty})}{1+\tilde{c}/(c_{+}/c_{\infty})}\right]$ is the sensing function, and $\Gamma = \ell_{\dd}/\ell_{\textrm{a}} = D_{c} k \rho_{\textrm{p}}/(v_0^2 c_{\textrm{M}})$ is the diffusio-absorption number defined in the Main Text. We impose the following boundary conditions:
\begin{subequations}\label{eq:bc}
\begin{gather}
\tilde{c} = 0, \quad \textrm{and} \quad \hat{\bm{x}} \cdot \bm{\nabla} \tilde{\rho} = 0, \quad \textrm{at} \, \, \tilde{x} = 0 \\
\tilde{c} = 1, \quad \textrm{and} \quad \hat{\bm{x}} \cdot \bm{\nabla} \tilde{\rho} = 0, \quad \textrm{at} \, \, \tilde{x} = \tilde L_x\\
\hat{\bm{y}} \cdot \bm{\nabla} \tilde{c} = 0 \quad \textrm{and} \quad \hat{\bm{y}} \cdot \bm{\nabla} \tilde{\rho} = 0 \quad \textrm{at} \, \, \tilde{y} = 0, \tilde L_y,
\end{gather}
\end{subequations}
where $\hat{\bm{x}}$ and $\hat{\bm{y}}$ are the Cartesian unitary vectors.

As an initial condition, we impose a cell density profile with a perturbation of dimensionless wavevector $\tilde{q} = 4\pi/\tilde L_y$ and amplitude $\tilde{A}$:
\begin{equation}\label{eq:initial}
\tilde{\rho}(\tilde{\bm{r}},\tilde{t} = 0) = \frac{1}{2}\left(1-\tanh \left[ \frac{\tilde{x}-1+\tilde{A} \cos(\tilde{q} \tilde{y})}{\tilde\delta} \right]\right),  
\end{equation}
where $\tilde\delta = 3$ sets the initial front width in the dimensionless simulation units. The initial chemoattractant concentration is determined by solving Eq.~\eqref{eq:bacterial} with the boundary conditions in \cref{eq:bc}.

We carry out numerical simulations using the finite-element method, which involves writing the equations in weak form by means of the integral scalar product using test functions $\tilde{\varphi}_{\rho}$ and $\tilde{\varphi}_c$ for the cell density $\tilde{\rho}(\tilde{\bm{r}},\tilde{t})$ and chemoattractant concentration $\tilde{c}(\tilde{\bm{r}},\tilde{t})$ fields, respectively. Using Green identities, we obtain an integral bilinear system of equations for the variables and their test functions:
\begin{subequations}
\begin{gather}
\begin{multlined}
\int_{\V} \dd\V \, \partial_{\tilde{t}} \tilde{c} \, \tilde{\varphi}_{c} + \frac{D_{\text{c}}}{\chi} \int_{\V} \dd \V \, \bm{\nabla} \tilde{c} \cdot \bm{\nabla} \tilde{\varphi}_{c} \\
- \frac{D_{\text{c}}}{\chi} \int_{\mathcal{C}} \dd \mathcal{C} \,\bm{\nabla} \tilde{c} \cdot \bm{n} \, \tilde{\varphi}_{c} + \frac{D_{\text{c}}}{\chi} \Gamma \int_{\V} \, \dd\V \,  \tilde{c} \tilde{\rho} \, \tilde{\varphi}_{c} = 0,
\end{multlined}\\
\begin{multlined}
 \int_{\V} \dd\V \, \partial_{\tilde{t}} \tilde{\rho} \, \tilde{\varphi}_{\rho} + \int_{\V} \dd\V \, \left[ \frac{D_\rho}{\chi}\bm{\nabla} \tilde{\rho} - \tilde{\rho} \bm{\nabla} f(\tilde{c}) \right] \cdot \bm{\nabla} \tilde{\varphi}_{\rho} \\ 
- \int_{\mathcal{C}} \dd \mathcal{C} \left[ \frac{D_\rho}{\chi} \bm{\nabla} \tilde{\rho} - \tilde{\rho} \bm{\nabla} f(\tilde{c})\right] \cdot \bm{n} \, \tilde{\varphi}_{\rho}  
 = 0.
\end{multlined}
\end{gather}
\end{subequations}
Here, $\bm{n}$ is the outer unit normal of the boundaries $\mathcal{C}$ of the 2D domain $\V$, and $\dd\V$ and $\dd\mathcal{C}$ are the surface and line elements, respectively. To ensure numerical stability, the equations are discretized in space using second-order Lagrange polynomials and triangular elements for the fields, and evolved in time through a 4th-order variable-step backward differentiation formula method. The relative tolerance of the nonlinear method is always set below $10^{-6}$. The time-dependent solver was complemented with an automatic refining mesh algorithm which increased the number of mesh elements depending on the absolute value of the cell density gradient at the traveling front.

\begin{figure*}[tb!]
\begin{center}
%\vskip-0.25cm
\includegraphics[width=0.9\textwidth]{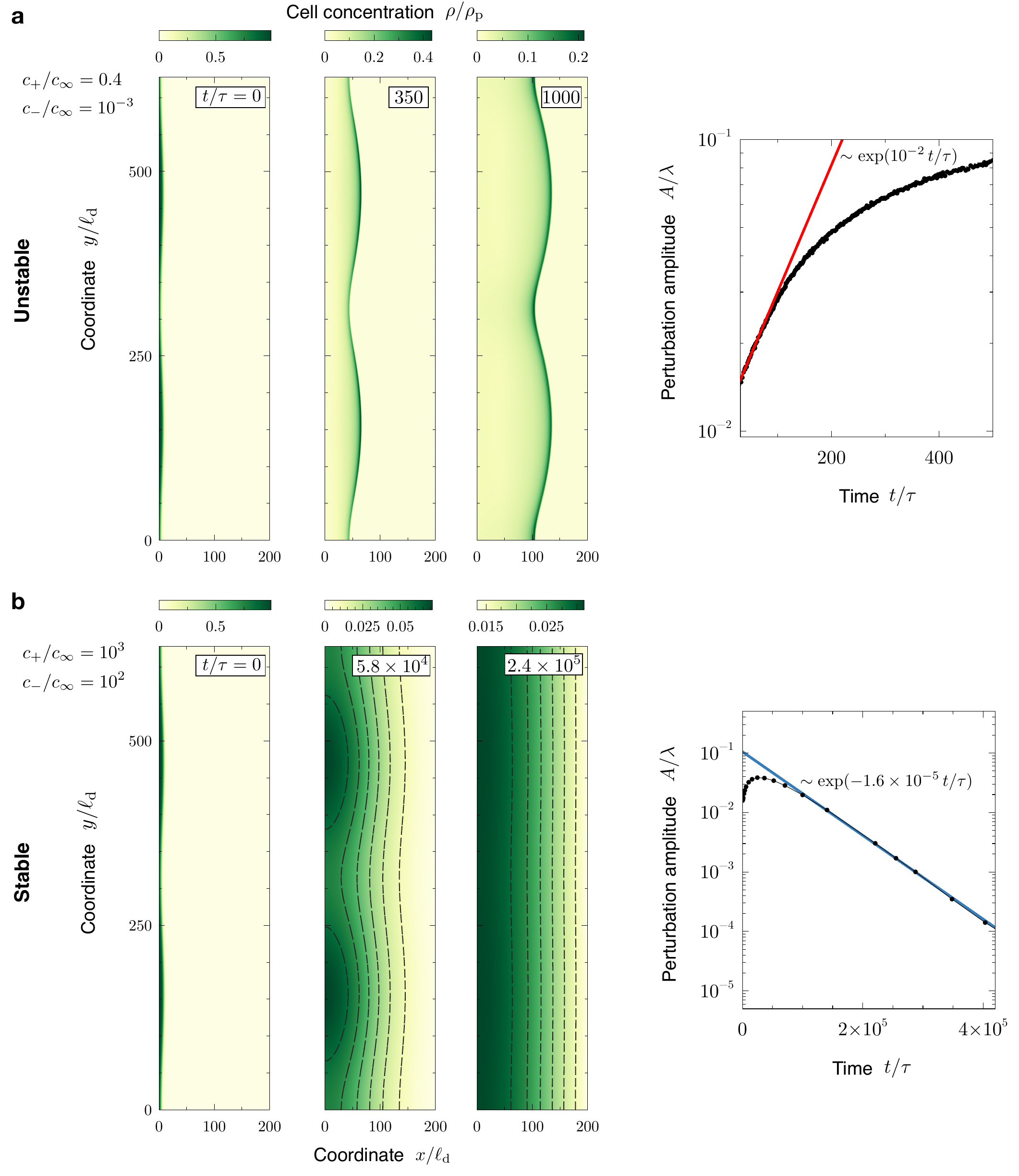}
\end{center}
  {\phantomsubcaption\label{Fig unstable}}
  {\phantomsubcaption\label{Fig stable}}
\bfcaption{Numerical simulations of stable and unstable chemotactic fronts}{ Snapshots of propagating cell pulses in the unstable (\subref*{Fig unstable}) and stable (\subref*{Fig stable}) regimes. In the stable case, dashed lines indicate isocontours of the cel concentration field. The cell pulse is initially perturbed as given by \cref{eq:initial} with a wavenumber $q\ell_{\text{d}} = 0.02$ and amplitude $A/\lambda = 0.016$, where $\lambda = 2\pi/q$. The right panels show the growth and decay of the perturbation amplitude over time for the unstable and stable cases, respectively. Red and blue lines are exponential fits that characterize the linear regime of the unstable and stable dynamics, respectively. The corresponding growth and decay rates are shown in the plots. Parameter values are $\Gamma = 11$, $D_\rho/\chi = 0.05$, and $D_{\text{c}}/\chi = 0.01$, and the unstable and stable regimes are obtained by choosing different values for the sensing concentrations $c_-$ and $c_+$ as indicated in the plot.} \label{Fig examples}
\end{figure*}

\begin{figure*}[tb!]
\begin{center}
%\vskip-0.25cm
\includegraphics[width=0.5\textwidth]{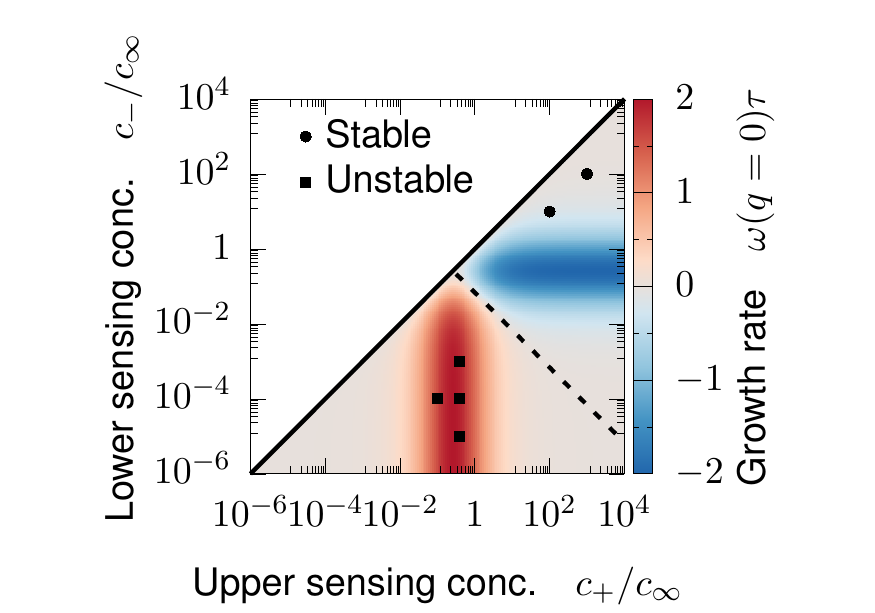}
\end{center}
\bfcaption{Simulation results (points) agree with the stability diagram predicted analytically}{ The dashed line indicates the stability limit given by \cref{eq stability-sensing}. Simulations with values of $c_-$ and $c_+$ falling within the stable and unstable regime indeed show stable and unstable front dynamics, respectively. Parameter values other than $c_-$ and $c_+$ are given in \cref{Fig examples}.} \label{Fig simulations-diagram}
\end{figure*}

\begin{figure*}[tb!]
\begin{center}
%\vskip-0.25cm
\includegraphics[width=0.75\textwidth]{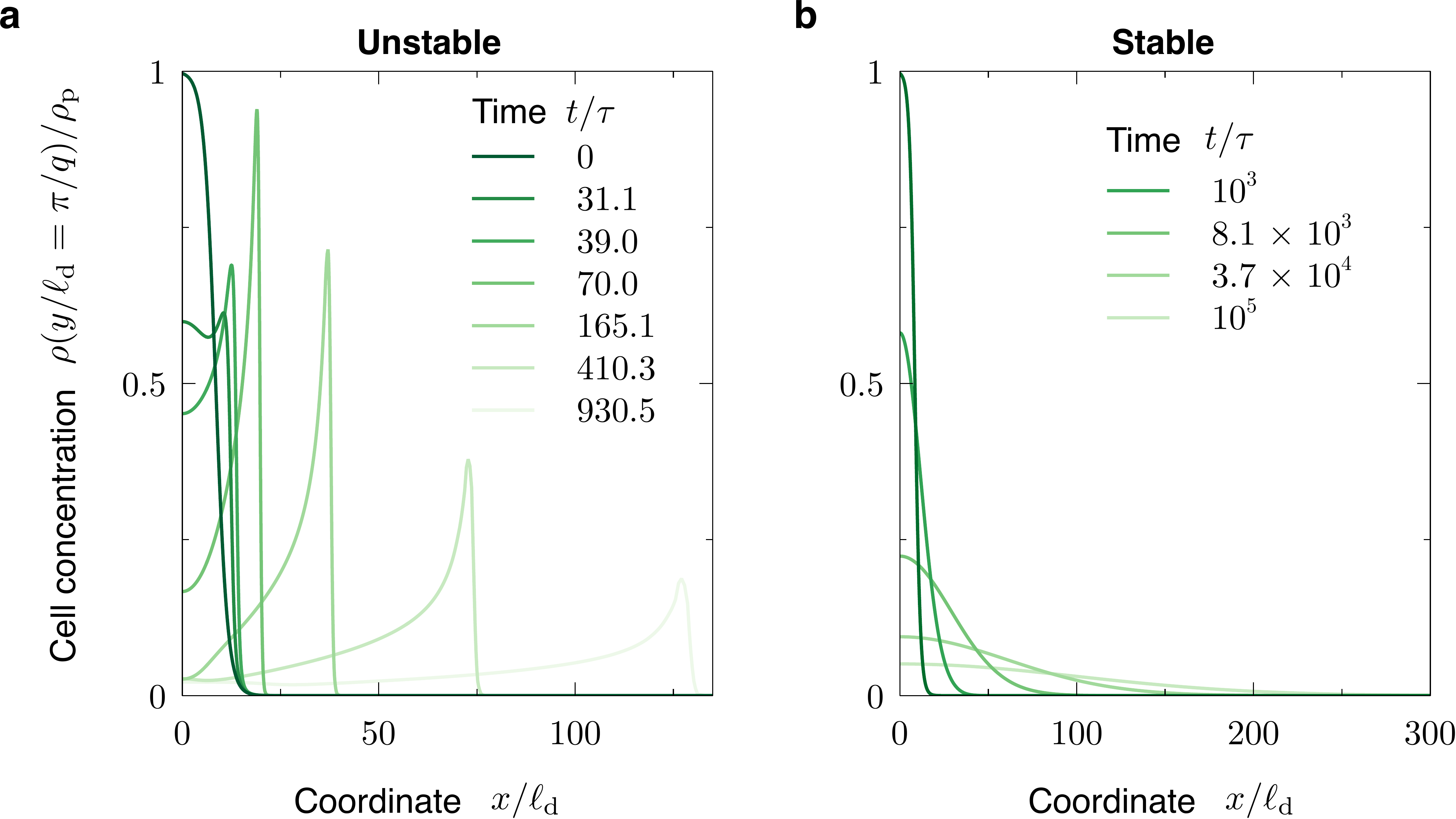}
\end{center}
  {\phantomsubcaption\label{Fig unstable-profiles}}
  {\phantomsubcaption\label{Fig stable-profiles}}
\bfcaption{Cell concentration profiles in numerical simulations}{ The cell concentration profile develops a sharper front in the unstable regime (\subref*{Fig unstable-profiles}) than in the stable one (\subref*{Fig stable-profiles}). Parameter values for each case are as given in \cref{Fig examples}.} \label{Fig profiles}
\end{figure*}

\onecolumngrid

\clearpage

\begin{table}[tb]
\begin{center}
\begin{tabular}{ll}

Description&Estimate\\\hline

Chemoattractant diffusivity&$D_{\text{c}}\sim 800$ $\mu$m$^2$/s\\
Maximal absorption rate per cell&$k\sim 2\times 10^6$ s$^{-1}$\\
%Maximal absorption rate&$k\sim 10^7$ s$^{-1}$\\
Half-maximum absorption concentr.&$c_{\text{M}}\sim 1$ $\mu$M\\
Far-field chemoattractant concentr.&$c_\infty\sim 10$ mM\\
Upper sensing concentration&$c_+\sim 30$ $\mu$M\\
Lower sensing concentration&$c_-\sim 1$ $\mu$M\\\hline
Effective cell diffusivity&$D_\rho\sim 0.9$ $\mu$m$^2$/s\\
Chemotactic susceptibility&$\chi\sim 9$ $\mu$m$^2$/s\\
Cell concentration in the front&$\rho_{\text{f}}\sim 0.0048$ $\mu$m$^{-3}$\\
Front speed&$v_0\sim 0.042$ $\mu$m/s\\\hline
Diffusion length&$\ell_{\text{d}} = D_{\text{c}}/v_0\sim 19$ mm\\
Absorption length&$\ell_{\text{a}} = v_0 c_{\text{M}}/(k\rho_{\text{f}})\sim 2.7$ nm\\
%Absorption length&$\ell_{\text{a}} = v_0 c_{\text{M}}/(k\rho_{\text{f}})\sim 0.5$ nm\\
Internal decay length&$\ell_{\text{i}} = \sqrt{\ell_{\text{d}} \ell_{\text{a}}}\sim 7$ $\mu$m\\
%Internal decay length&$\ell_{\text{i}} = \sqrt{\ell_{\text{d}} \ell_{\text{a}}}\sim 3$ $\mu$m\\
Diffusio-absorption number&$\Gamma = \ell_{\text{d}}/\ell_{\text{a}} \sim 7.2\times 10^6$\\
%Diffusio-absorption number&$\Gamma = \ell_{\text{d}}/\ell_{\text{a}} \sim 3.5\times 10^7$\\
Chemotactic response number&$\alpha = f'_0 c_\infty \sim 1.8\times 10^3$\\
%Chemotactic response number&$\alpha = f'_0 c_\infty \sim 3.4\times 10^3$\\
Response limitation number&$\beta = f''_0 c_\infty^2 \sim 4.4\times 10^6$
%Response limitation number&$\beta = f''_0 c_\infty^2 \sim 1.4\times 10^7$

\end{tabular}

\caption{Estimates of model parameters. The values correspond to \textit{E. coli} cells migrating toward the amino acid serine through porous media, as in Ref. \cite{Bhattacharjee2021a}. The first and second parts of the table correspond to parameters related to chemoattractant and cell motion, respectively. The third part corresponds to parameters derived here from the above. To obtain $\alpha$ and $\beta$, we evaluate $f'_0 = f'(c_0(0))$ and $f''_0= f''(c_0(0))$ using \cref{eq chemo-flat-solutions} to calculate $c_0(0)$.} \label{t parameters}
\end{center}
\end{table}

\begin{table}[tb!]
\begin{center}
%\vskip-0.5cm
\begin{tabular}{l|ccc|ccc}

&\multicolumn{6}{c}{\textbf{Experiment}}\\
&\multicolumn{3}{c|}{\textit{E. coli} and aspartate, Fu et al. \cite{Fu2018}}&\multicolumn{3}{c}{\textit{E. coli} and serine, Bhattacharjee et al. \cite{Bhattacharjee2021a}}\\
\textbf{Parameter}&$c_\infty = 50$ $\mu$M&$c_\infty = 100$ $\mu$M&$c_\infty = 200$ $\mu$M&$\xi = 1.2$ $\mu$m&$\xi = 1.7$ $\mu$m&$\xi = 2.2$ $\mu$m\\\hline

Chemoattractant diffusivity $D_{\text{c}}$ ($\mu$m$^2$/s)&$500$&$500$&$500$&$800$&$800$&$800$\\
Maximal absorption rate per cell $k$ (s$^{-1}$)&$9.3\times 10^4$&$9.3\times 10^4$&$9.3\times 10^4$&$9.6\times 10^6$&$9.6\times 10^6$&$9.6\times 10^6$\\
Half-maximum absorption concentr. $c_{\text{M}}$ ($\mu$M)&$0.5$&$0.5$&$0.5$&$1$&$1$&$1$\\
Far-field chemoattractant concentr. $c_\infty$ ($\mu$M)&$50$&$100$&$200$&$10^4$&$10^4$&$10^4$\\
Upper sensing concentration $c_+$ ($\mu$M)&$10^3$&$10^3$&$10^3$&$30$&$30$&$30$\\
Lower sensing concentration $c_-$ ($\mu$M)&$3.5$&$3.5$&$3.5$&$1$&$1$&$1$\\\hline
Effective cell diffusivity $D_\rho$ ($\mu$m$^2$/s)&$165$&$165$&$165$&$0.4$&$0.9$&$2.3$\\
Chemotactic susceptibility $\chi$ ($\mu$m$^2$/s)&$3.6\times 10^3$&$3.6\times 10^3$&$3.6\times 10^3$&$5$&$9$&$145$\\
Cell concentration in the pulse $\rho_{\text{p}}$ ($\mu$m$^{-3}$)&$0.0015$&$0.0038$&$0.010$&$0.014$&$0.0048$&$0.048$\\
Front speed $v_0$ ($\mu$m/s)&$5.0$&$3.4$&$3.2$&$0.017$&$0.042$&$0.25$\\\hline
Diffusion length $\ell_{\text{d}} = D_{\text{c}}/v_0$ (mm)&$0.10$&$0.15$&$0.16$&$47$&$19$&$3.2$\\
Absorption length $\ell_{\text{a}} = v_0 c_{\text{M}}/(k\rho_{\text{p}})$ ($\mu$m)&$13$&$3.4$&$1.1$&$7.6\times 10^{-5}$&$5.5\times 10^{-4}$&$3.3\times 10^{-4}$\\
Internal decay length $\ell_{\text{i}} = \sqrt{\ell_{\text{d}} \ell_{\text{a}}}$ ($\mu$m)&$36$&$23$&$13$&$1.9$&$3.2$&$1.0$\\
Diffusio-absorption number $\Gamma = \ell_{\text{d}}/\ell_{\text{a}}$&$11$&$57$&$161$&$6.2\times 10^8$&$3.5\times 10^7$&$9.8\times 10^6$\\
Chemotactic response number $\alpha = f'_0 c_\infty$&$2.8$&$5.9$&$10$&$6.8\times 10^3$&$3.4\times 10^3$&$2.1\times 10^3$\\
Response limitation number $\beta = f''_0 c_\infty^2$&$8.2$&$36$&$110$&$5.1\times 10^7$&$1.4\times 10^7$&$5.6\times 10^6$

\end{tabular}

\caption{Estimates of parameter values for experiments of bacterial chemotactic fronts. In the experiments by Fu et al. \cite{Fu2018}, bacteria swim in liquid media with three different initial chemoattractant concentrations $c_\infty$. In the experiments by Bhattacharjee et al. \cite{Bhattacharjee2021a}, bacteria swim through porous media of three different pore sizes $\xi$. As in \cref{t parameters}, the first and second parts of the table correspond to parameters related to chemoattractant and cell motion, respectively. The third part corresponds to parameters derived here from the above. To obtain $\alpha$ and $\beta$, we evaluate $f'_0 = f'(c_0(0))$ and $f''_0= f''(c_0(0))$ using \cref{eq chemo-flat-solutions} to calculate $c_0(0)$.} \label{t sensing-estimates}
\end{center}
\end{table}

\begin{figure}[tb!]
\begin{center}
%\vskip-0.25cm
\includegraphics[width=\textwidth]{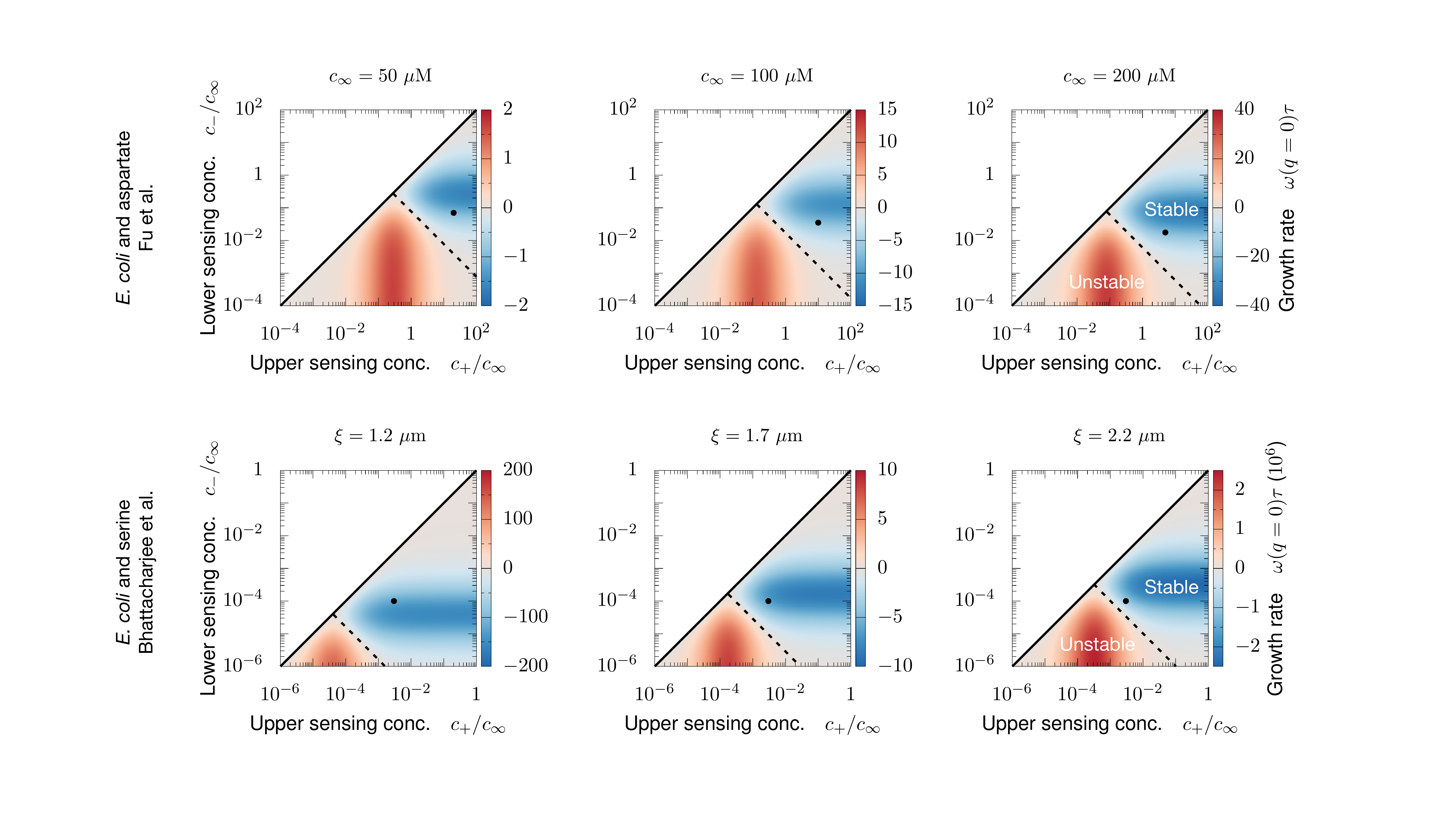}
\end{center}
\bfcaption{Stability of chemotactic fronts of \textit{E. coli} in experiments}{ In the experiments by Fu et al. \cite{Fu2018}, \textit{E. coli} swim in liquid media with three different initial concentrations $c_\infty$ of the chemoattractant aspartate. In the experiments by Bhattacharjee et al. \cite{Bhattacharjee2021a}, \textit{E. coli} swim through porous media of three different pore sizes $\xi$. For each experiment, the stability diagram is plotted using the parameter values in \cref{t sensing-estimates}. The points correspond to the actual experimental conditions, namely the values of $c_+/c_\infty$ and $c_-/c_\infty$ in each case. In all cases, our analysis predicts stable fronts, consistent with the experimental observation of flat fronts.} \label{Fig experiment}
\end{figure}

\end{document}